\documentclass{llncs}

\usepackage{polski}
\usepackage[utf8]{inputenc}

\usepackage{graphicx}
\usepackage[utf8]{inputenc}
\usepackage{listings}
\usepackage{balance}
\usepackage{url}
\usepackage{tabularx}
\usepackage{csquotes}
\usepackage{enumitem}
\usepackage{multirow}
\usepackage{float}
\usepackage{todonotes}

\usepackage[misc]{ifsym}

\usepackage{fancyhdr} 

\usepackage{ifthen}

\begin{document}

\title {Defect prediction with bad smells in code}
\author{Jarosław Hryszko\inst{1,2}
\and 
Lech Madeyski\inst{1} % (\Letter)
\and
Marta Dąbrowska\inst{1,2}
\and Piotr Konopka\inst{1,2}
}

\institute{Faculty of Computer Science and Management\\
Wroclaw University of Science and Technology
\and
Volvo Group
}

\maketitle
\thispagestyle{fancy}
\rhead{\ifthenelse{\value{page}=1}{\tiny{Jarosław Hryszko, Lech Madeyski, Marta Dąbrowska and Piotr Konopka, “\textit{Defect Prediction with Bad Smells in Code}” in Software Engineering: Improving Practice through Research (B. Hnatkowska and M. Śmiałek, eds.), pp. 163–176, 2016. URL: http://madeyski.e-informatyka.pl/download/Hryszko16.pdf}}{}}

\begin{abstract}
\emph{Background.} Defect prediction in software can be highly beneficial for development projects, when prediction is highly effective and defect-prone areas are predicted correctly. One of the key elements to gain effective software defect prediction is proper selection of metrics used for dataset preparation. 

\emph{Objective.} The purpose of this research is to verify, whether code smells metrics, collected using Microsoft CodeAnalysis tool, added to basic metric set, can improve defect prediction in industrial software development project.

\emph{Results.} We verified, if dataset extension by the code smells sourced metrics, change the effectiveness of the defect prediction by comparing prediction results for datasets with and without code smells-oriented metrics. In a result, we observed only small improvement of effectiveness of defect prediction when dataset extended with bad smells metrics was used: average accuracy value increased by 0.0091 and stayed within the margin of error. However, when only use of code smells based metrics were used for prediction (without basic set of metrics), such process resulted with surprisingly high accuracy (0.8249) and F-measure (0.8286) results. We also elaborated data anomalies and problems we observed when two different metric sources were used to prepare one, consistent set of data.

\emph{Conclusion.} Extending the dataset by the code smells sourced metric does not significantly improve the prediction effectiveness. Achieved result did not compensate effort needed to collect additional metrics. However, we observed that defect prediction based on the code smells only is still highly effective and can be used especially where other metrics hardly be used.
\end{abstract}

\section{Introduction}
Among different aspects of software defect prediction process, one of the key elements is proper selection of metrics for training and verification dataset preparation. Most popular data is source code metrics~\cite{Hall2012,Nam2014}, but also different types of metrics are considered effective in term of defect prediction, such as design metrics~\cite{Succi}, change metrics~\cite{Moser2008}, mining metrics~\cite{Nagappan2006} or process metrics~\cite{Madeyski15SQJ,Madeyski11}.

\subsection{Related work and goal}
Separate group of design metrics are metrics based on code smells, also known as bad smells or code bad smells. The term was formulated by Kent Beck in 2006~\cite{CodeSmell}. The concept was popularized by Martin Fowler in his book \emph{Refactoring. Improving the structure of existing code}~\cite{FowlerREF2006}. Kent Beck was a co-author of the chapter on code smells.

Kent Beck on his website explains the idea of code smells:
\begin{displayquote}
Note that a Code Smell is a hint that something might be wrong, not a certainty. A perfectly good idiom may be considered a Code Smell because it's often misused, or because there's a simpler alternative that works in most cases. Calling something a Code Smell is not an attack; it's simply a sign that a closer look is warranted.~\cite{CodeSmell}
\end{displayquote}

Due to nature of code smells described above, there is ongoing discussion if code smells could be used effectively in quality assurance in code development~\cite{ZhangTROUBLE2008,Zhang2011}. Major motivation for this research was to investigate, if code smells can improve software defect prediction.

In industrial software development, only Holschuh et al. investigated code smells metrics effectiveness in defect prediction process for Java programming language~\cite{Holschuh2009}. No code smells metrics for defect prediction in .NET oriented industrial software projects are known to authors. Thus, we decided use long-term defect prediction research project run in Volvo Group~\cite{HryszkoASDP2016,HryszkoADPD2016} as an occasion for conducting an experiment with introduction of bad smells based metrics to prediction process and observe the results, if they improved prediction effectiveness or not:

\textbf{RQ:} How Code Bad Smells based metrics impact defect prediction in industrial software development project? 

\subsection{Research environment: Industrial software development project}
Project, on which the study was conducted, is a software development of critical industry system used in Volvo Group vehicle factories called PROSIT+. It is created based on client-server architecture. The main functionality of PROSIT+ system is: programming, testing, calibration and electrical assembly verification of Electronic Control Units (ECUs) in Volvo's vehicle production process.

PROSIT+ system consists of few coexisting applications. The most important one, desktop application -- "PROSIT Operator", communicates in real time with a mobile application, located on palmtop computer used by vehicle factory workers to transfer all production related information to a local server. The server is responsible for storage and distribution of configuration-, system- and product-related data. Such communication can generate extremely heavy data transfer loads in large factories, when more than 100 mobile applications are used.
Other application include: "PROSIT Designer", "PROSIT Factory Manager" and web application "PROSIT Viewer". All of them are also connected to the same server.

Development of each PROSIT+ version lasts one year. After this period software is released to the end-user. As this period of time is connected to factory production cycle it cannot be fastened or postponed.

All applications within PROSIT+ system were developed using Microsoft .NET technology and Microsoft Visual Studio as the integrated development environment. For version control purposes, Microsoft Team Foundation Server was used. Before release of version 11 of the PROSIT+ system, IBM ClearQuest was used for software defect management. Until the development of version 11, Team Foundation Server was used for defect tracking.

Project lacks of bottlenecks described by Hryszko and Madeyski\cite{Hryszko2015}, which could hinder or prevent from applying defect prediction process. However, we observed relatively high number of naming issues in the project. Main reason of that situation we consider high maturity of the software system -- over the time, naming conventions have changed. We consider naming issues as negligible problem and we will exclude them from the further investigation.

\section{Research process}
Defect prediction was already an ongoing process in investigated project. It used SourceMonitor software as metric source and as prediction tool -- KNIME-based DePress Extensible Framework proposed by Madeyski and Majchrzak~\cite{MadeyskiDEP2014}. This tool, based on KNIME~\cite{KNIMEDocumentation}, provides with a wide range of data-mining techniques, including defects prediction, in various IT projects, independently of technology and programming language used. We will also use KNIME/DePress for purpose of our research.

To investigate the possible impact of code-smell metrics on defect prediction, we developed the following plan to follow:
\begin{enumerate}
	\item Generate metrics from SourceMonitor;
	\item Generate code smells metrics from CodeAnalysis;
	\item Parse results from CodeAnalysis and merge them with metrics from SourceMonitor.
	\item Link check-ins to defects;
	\item Link classes from check-ins to defects (the assumption is that if a class was changed while fixing a defect, that class was partially or fully responsible for that defect);
	\item Merge list of classes with merged metrics from CodeAnalysis and SourceMonitor;
	\item Use different software defect prediction approaches combinations to select optimal prediction set-up for evaluation purposes;
	\item Divide PROSIT+ code into 20 sub-modules and run prediction model training and evaluation using data from each module separately;
	\item Collect and interpret the results. 
\end{enumerate}
	
\subsection{SourceMonitor as basic metrics source}
Defect prediction process in PROSIT+ is based on metrics that are gathered using SourceMonitor tool~\cite{SMSite}. That tool performs static computer code analysis on complete files and extracts 24 different kinds of metrics. Example metrics extracted are:
\begin{itemize}
	\item Lines of code,
	\item Methods per class,
	\item Percentage of comments,
	\item Maximum Block Depth,
	\item Average Block Depth.
\end{itemize}

\subsection{CodeAnalysis tool as code smells metrics source}
In our experiment, we decided to use Microsoft CodeAnalysis tool to gather code smells metrics. Primary deciding factor was cost: CodeAnalysis tool is delivered as a part of Microsoft Visual Studio software development suite for .NET based projects. Thus, there was no additional costs of introduction of this tool into the investigated software development project. 

\begin{displayquote}
    CodeAnalysis for managed code analyzes managed assemblies and reports information about the assemblies, such as violations of the programming and design rules set forth in the Microsoft .NET Framework Design Guidelines~\cite{MicrosoftCA}.
\end{displayquote}

According to documentation, there are approximately two hundred rules in CodeAnalysis~\cite{MicrosoftCA}, trigerring 11 kinds of warnings (Table~\ref{table:warnings}). Tool can be run from command line and results are then stored in an .xml file, that can be later parsed and analyzed further.
	
\begin{table}
\caption{Bad smell warnings in CodeAnalysis}
\label{table:warnings}
		\begin{tabular}{| l | p{9cm} |}
			\hline
			Bad smell warning & Area covered\\ \hline
			Design & Correct library design as specified by the .NET Framework Design Guidelines\\ \hline
			Globalization & World-ready libraries and applications\\ \hline
			Interoperability & Interaction with COM clients\\ \hline
			Maintainability & Library and application maintenance\\ \hline
			Mobility & Efficient power usage\\ \hline
			Naming & Adherence to the naming conventions of the .NET Framework Design Guidelines\\ \hline
			Performance & High-performance libraries and applications\\ \hline
			Portability & Portability across different platforms\\ \hline
			Reliability & Library and application reliability, such as correct memory and thread usage\\ \hline
			Security & Safer libraries and applications\\ \hline
			Usage & Appropriate usage of the .NET Framework\\
			\hline
		\end{tabular}
\end{table}

\section{Results}
We conducted our experiment by following the plan presented in previous section. Here we present the results.

\subsection{Automatically generated code: observed anomaly, cause and solution}	\label{sec:anomaly}
\begin{figure}
	\includegraphics[width=\linewidth]{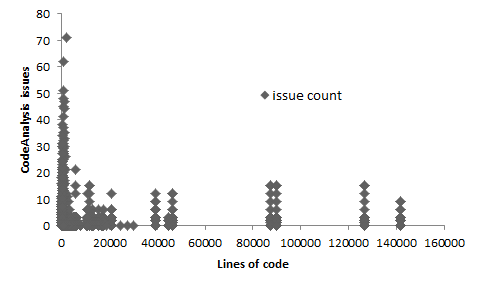}
	\caption{Anomalies in number of issues metric per file length (measured in LOC) relation, introduced by automatically generated code, later removed from analysis}
	\label{fig:CAautomated}
\end{figure}

After analyzing the relation between number of reported code smells issues and file length metrics for complete software system, in datasets prepared basing on CodeAnalysis and SourceMonitor tools, we observed that different number of issues are reported for the same, large file length values (Figure~\ref{fig:CAautomated}). As considered software contains only small number of large files, we interpreted that as an anomaly: different total number of code bad smell issues were reported for the same files. After investigation, we found that in investigated system files with more than 1000 lines of code (LOC) are in most cases generated automatically and contain more than one class for a file, while CodeAnalysis tool calculates number of issues metric per class. That discrepancy resulted in abnormal number of issue per file length relation: different \emph{number of issues} values were collected for the same \emph{LOC} values, because \emph{number of issues} values were calculated for different classes located in the same files, identified by the same \emph{LOC} value.

\begin{figure}
	\includegraphics[width=\linewidth]{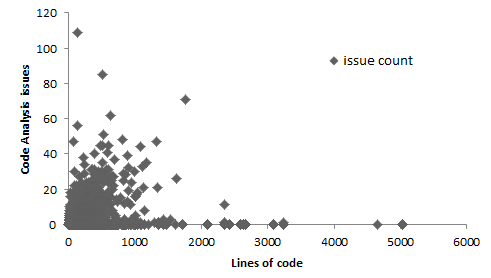}
	\caption{Number of issues metric per file length (measured in LOC) relation for investigated software, with automatically generated code removed}
	\label{fig:CAnotautomated}
\end{figure}

As automatically generated code files exist only for installation and deployment purposes and are not covered by tests and are not reachable for end-users of the system, we decided to consider them as a source of information noise and \textbf{we removed them from further analysis}. Number of issue per file length relation improved after that step (Figure~\ref{fig:CAnotautomated}).

\subsection{Metrics breakdown difference: problem and solution}
After a thorough investigation of the above problem, we found that different values of issue number metric for the same LOC metric was caused by the different metrics breakdown used by two tools selected for metric datasets generation: CodeAnalysis gathers data for every class while SourceMonitor for every file. When results from two tools were merged into single dataset, SourceMonitor metrics, fixed for each file, were artificially divided per each class in the file (Table~\ref{table:Example1}). 

To counteract against metric anomalies described in section~\ref{sec:anomaly}, as well as against possible introduction of informational noise into the training dataset, we decided to change the approach and rearrange the datasets into single file metrics per record layout. To achieve this, metrics gathered by CodeAnalysis had to be aggregated (added; Table~\ref{table:Example2}).

\begin{table}
\begin{center}
\label{table:Example1}
\caption{Example of dataset from first approach: single class per record (SourceMonitor metrics are artifically divided per each class in file)}
\begin{tabular}{| p{2cm} | p{2cm} | p{3cm} | p{3cm} |}
\hline
File & Class & SourceMonitor LOC & CodeAnalysis Issues \\ \hline
File1.cs & Class1 & 33& 3\\ \hline
File1.cs & Class2 & 33& 20 \\ \hline
File1.cs & Class3 & 33& 6\\ \hline
File2.cs & Class4 & 30& 15 \\
\hline
\end{tabular}
\end{center}
\end{table}

\begin{table}
\begin{center}
\caption{Example of dataset from second approach: single file per record (CodeAnalysis metrics are artificially added)}
\label{table:Example2}
\begin{tabular}{| p{2cm} | p{2cm} | p{3cm} | p{3cm} |}
\hline
File & Class & SourceMonitor LOC & CodeAnalysis Issues \\ \hline
File1.cs & Class1...3 & 100& 29 \\ \hline
File2.cs & Class4 & 30& 15 \\
\hline
\end{tabular}
\end{center}
\end{table}

\subsection{Optimal prediction mechanism selection}
To choose optimal prediction mechanism, we decided to test combination of different classifiers, feature selection and balance algorithms (Table~\ref{table:Factors}) against two datasets: with- and without code bad smells metrics collected by CodeAnalysis tool.

We used SMOTE algorithm~\cite{ChawlaSMOTE2002} to balance classes with defects and without them.

To select most important metrics from all available, as some of them should have seemingly little impact on the presence of true software defects, e.g. \emph{Efficient power usage} warning (Table~\ref{table:warnings}), we decided to use in our research two feature selection algorithms: KNIME's build-in reversed elimination greedy algorithm~\cite{KNIMEElimination} and simulated annealing meta-heuristic algorithm by Kirkpatrick et al.~\cite{KirkpatrickANN1983} in form proposed by Brownlee~\cite{BrownleeALG2011}.

As classifier, we used popular in defect prediction studies~\cite{Hall2012,Moser2008,Khosh1995,Selby1988} Naive Bayes classifier and Probabilistic Neural Network (PNN), as well as Random Forest~\cite{BreimanRF2001} classifier.

Results of testing combinations of above machine learning elements in favor of best prediction results are presented in Table~\ref{table:Results}. Two datasets -- with- and without code bad smells metrics included, were divided using stratified sampling method into two equal subsets, for training and evaluation purpose. Prediction models were evaluated using F-measure~\cite{Witten2005}.

Highest F-measure value (0.9713) was observed for dataset with code bad smells used, when SMOTE algorithm and reversed elimination feature selection mechanism was used to select optimal subset for training and evaluation of Random Forest classifier. And such combination was selected for final evaluation of usage of code smells based metrics in defect prediction process.

\begin{table}
\begin{center}
\caption{Combinations of different approaches}
\label{table:Factors}
\begin{tabular}{| p{3cm} | p{3cm} | p{2cm} | p{3cm} |}
\hline
Classifier& Feature Selection& SMOTE& Bad smells metrics? \\ \hline
Naive Bayes& None& With& Present \\ 
Random Forest & Elimination & Without & Absent\\ 
PNN& Simulated Annealing & ~& ~ \\
\hline
\end{tabular}
\end{center}
\end{table}
	
\begin{table}
\caption{Results for optimal prediction set-up selection (defect-prone class)}
\label{table:Results}
    \resizebox{\textwidth}{!}{
    \begin{tabular}{|c|c|l|l|l|l|l|l|l|l|l|l|l|}
        \hline
        \multicolumn{1}{|c|}{\multirow{3}{*}{Classifier}} &
        \multicolumn{1}{c|}{\multirow{3}{*}{SMOTE}} &
        \multicolumn{1}{c|}{\multirow{3}{*}{\vtop{\hbox{\strut Feature}\hbox{\strut selection}}}} &
        \multicolumn{10}{c|}{Bad smells metrics included?} \\
        \cline{4-13} 
        \multicolumn{1}{|c|}{} &
        \multicolumn{1}{c|}{} &
        \multicolumn{1}{c|}{} &
        \multicolumn{5}{c|}{No} &
        \multicolumn{5}{c|}{Yes} \\
        \cline{4-13} 
        \multicolumn{1}{|c|}{} &
        \multicolumn{1}{c|}{} &
        \multicolumn{1}{c|}{} &
        F-meas.& TP& FP & TN& FN& F-meas.& TP& FP & TN& FN\\
        \hline
        \multirow{6}{*}{\vtop{\hbox{\strut Naive}\hbox{\strut Bayes}}} &
        \multirow{3}{*}{NO}&
        Annealing & 0.1318 & 23& 161 & 3330 & 142 & 0.1149 & 15& 81 & 3410 & 150 \\ 
        \cline{3-13} 
        & \multicolumn{1}{c|}{} & Elimination & 0 & 0& 1& 3490 & 165 & 0 & 0& 0& 3491 & 165 \\ 
        \cline{3-13} 
        & \multicolumn{1}{c|}{} & None & 0.1314 & 31& 276 & 3215 & 134 & 0.1389 & 40& 371 & 3120 & 125 \\ 
        \cline{2-13} 
        & \multirow{3}{*}{YES} & Annealing & 0.5474 & 1497 & 482 & 3008 & 1994 & 0.586 & 1695 & 599 & 2891 & 1796 \\ 
        \cline{3-13} 
        & & Elimination & 0.5961 & 1736 & 598 & 2892 & 1755 & 0.6106 & 1807 & 621 & 2869 & 1684 \\ 
        \cline{3-13} 
        & & None & 0.5424 & 1476 & 475 & 3015 & 2015 & 0.5775 & 1657 & 591 & 2899 & 1834 \\ 
        \hline
        \multirow{6}{*}{PNN} &
        \multirow{3}{*}{NO} & Annealing & 0 & 0& 0& 3491 & 165 & 0 & 0& 0& 3491 & 165 \\ 
        \cline{3-13} 
        & & Elimination & 0 & 0& 0& 3491 & 165 & 0 & 0& 0& 3491 & 165 \\ 
        \cline{3-13} 
        & & None & 0 & 0& 0& 3491 & 165 & 0 & 0& 0& 3491 & 165 \\ 
        \cline{2-13} &
        \multirow{3}{*}{YES} & Annealing & 0.7313 & 2187 & 303 & 3187 & 1304 & 0.7582 & 2335 & 333 & 3157 & 1156 \\ 
        \cline{3-13} 
        & & Elimination & 0 & 0& 0& 3491 & 165 & 0.8051 & 2568 & 320 & 3170 & 923 \\ 
        \cline{3-13} 
        & & None & 0.7253 & 2147 & 282 & 3208 & 1344 & 0.7333 & 2204 & 316 & 3174 & 1287 \\ 
        \hline
        \multirow{6}{*}{\vtop{\hbox{\strut Random}\hbox{\strut Forest}}} & 
        \multirow{3}{*}{NO} & Annealing & 0.0963 & 9& 13 & 3478 & 156 & 0.1538 & 15& 15 & 3476 & 150 \\ 
        \cline{3-13} 
        & & Elimination & 0.1587 & 15& 9& 3482 & 150 & 0.1405 & 13& 7& 3484 & 152 \\ 
        \cline{3-13} 
        & & None & 0.1429 & 14& 17 & 3474 & 151 & 0.1538 & 15& 15 & 3476 & 150 \\ 
        \cline{2-13} & 
        \multirow{3}{*}{YES} & Annealing & 0.9551 & 3364 & 189 & 3301 & 127 & 0.9696 & 3407 & 130 & 3360 & 84\\ 
        \cline{3-13} 
        & & Elimination & 0.9654 & 3390 & 142 & 3348 & 101 & \textbf{0.9713} & 3435 & 147 & 3343 & 56\\ 
        \cline{3-13} 
        & & None & 0.9601 & 3383 & 173 & 3317 & 108 & 0.9696 & 3407 & 130 & 3360 & 84\\ 
        \hline
    \end{tabular}}
\end{table}

\subsection{Datasets evaluation: CodeAnalysis (bad smells metrics) against SourceMonitor}
For final evaluation, if code bad smells-based metrics could be valuable for defect prediction purposes, we divided all available code, in considered industrial software development project, into 20 smaller, similar in size sub-modules (ca. 700 records after SMOTE oversampling). Greater fragmentation of system's code was not technically possible. For each sub-module we collected metrics using SourceMonitor or/and CodeAnalysis, to create different datasets:
\begin{itemize}
    \item 20 datasets of SourceMonitor metrics only;
    \item 20 datasets of CodeAnalysis (code smells) metrics only;
    \item 20 datasets of combined metric: SourceMonitor + CodeAnalysis.
\end{itemize}
Additionally, each kind of datasets we decided to test against feature selection (FS) process. During the evaluation, we collected Accuracy and Cohen's kappa measures for overall results (Table~\ref{table:a_stat_table}), and F-measure and Recall for defect-prone classes (Table~\ref{table:f_stat_table}).

\begin{table}
\centering
\caption{Final results of datasets evaluation}
\label{table:a_stat_table}
\begin{tabular}{|l|l|l|c|}
\hline
Dataset                                 & Measure        & Mean   & Std. deviation \\ \hline
\multirow{2}{*}{SourceMonitor without FS}       & Accuracy      & 0.9422 & 0.0187         \\ \cline{2-4} 
                                      & Cohen's kappa & 0.8844 & 0.0374         \\ \hline
\multirow{2}{*}{CodeAnalysis without FS}          & Accuracy      & 0.676  & 0.0451         \\ \cline{2-4} 
                                      & Cohen's kappa & 0.3518 & 0.0904         \\ \hline
\multirow{2}{*}{SourceMonitor + CodeAnalysis w/o FS} & Accuracy      & 0.9487 & 0.0226         \\ \cline{2-4} 
                                      & Cohen's kappa & 0.8973 & 0.0453         \\ \hline
\multirow{2}{*}{SourceMonitor with FS}         & Accuracy      & 0.97   & 0.0122         \\ \cline{2-4} 
                                      & Cohen's kappa & 0.9399 & 0.0245         \\ \hline
\multirow{2}{*}{CodeAnalysis with FS}            & Accuracy      & 0.8249 & 0.059          \\ \cline{2-4} 
                                      & Cohen's kappa & 0.6497 & 0.1180         \\ \hline
\multirow{2}{*}{SourceMonitor + CodeAnalysis with FS}      & Accuracy      & \textbf{0.9791} & 0.0135         \\ \cline{2-4} 
                                      & Cohen's kappa & 0.9582 & 0.027          \\ \hline
\end{tabular}
\end{table}

\begin{table}
\centering
\caption{Measures for records marked as defect-prone}
\label{table:f_stat_table}
\begin{tabular}{|l|l|l|c|}
\hline
Dataset                                 & Measure                           & Mean    & Std. deviation \\ \hline
\multirow{2}{*}{SourceMonitor without FS}       & Recall                           & 0.9608  & 0.0278         \\ \cline{2-4} 
                                      & F-measure                        & 0.9433  & 0.0188         \\ \cline{2-4} 
                                      %& Matthews correlation coefficient & 0.0832  & 0.0809         \\ 
                                      \hline
\multirow{2}{*}{CodeAnalysis without FS}          & Recall                           & 0.666   & 0.2961         \\ \cline{2-4} 
                                      & F-measure                        & 0.6447  & 0.1157         \\ \cline{2-4} 
                                      %& Matthews correlation coefficient & -0.0313 & 0.658          \\ 
                                      \hline
\multirow{2}{*}{SourceMonitor + CodeAnalysis w/o FS} & Recall                           & 0.9637  & 0.0303         \\ \cline{2-4} 
                                      & F-measure                        & 0.9494  & 0.0228         \\ \cline{2-4} 
                                      %& Matthews correlation coefficient & 0.0733  & 0.0733         \\ 
                                      \hline
\multirow{2}{*}{SourceMonitor with FS}         & Recall                           & 0.9824  & 0.0146         \\ \cline{2-4} 
                                      & F-measure                        & 0.9704  & 0.012          \\ \cline{2-4} 
                                      %& Matthews correlation coefficient & 0.0725  & 0.066          \\ 
                                      \hline
\multirow{2}{*}{CodeAnalysis with FS}            & Recall                           & 0.8424  & 0.0542         \\ \cline{2-4} 
                                      & F-measure                        & 0.8286  & 0.0559         \\ \cline{2-4} 
                                      %& Matthews correlation coefficient & 0.0446  & 0.0704         \\ 
                                      \hline
\multirow{2}{*}{SourceMonitor + CodeAnalysis with FS}      & Recall                           & 0.9859  & 0.0206         \\ \cline{2-4} 
                                      & F-measure                        & 0.9792  & 0.0136         \\ \cline{2-4} 
                                      %& Matthews correlation coefficient & 0.0579  & 0.0742         \\ 
                                      \hline
\end{tabular}
\end{table}

\subsection{Threads to validity}
\emph{Conclusion validity.} In our research, we tested 20 datasets collected from different software modules. More research using larger data set, collected from different sources is needed to confirm our findings.

\emph{Internal validity.} We have used aggregation of CodeAnalysis metrics for each file, by adding metrics collected for each class. Such solution was introduced to solve metrics breakdown difference problem and make combination of two metric sources possible, however it could impact the final result of our research.

\emph{External validity.} Our research is based only on metrics gathered from one software development project. Despite the fact, that we were able to collect 34 different metric kinds for 20 different program modules, we were still constrained by single environment: development team and its programming habits, programming language, tools used, etc. Because of this fact, more research is needed to verify our findings in other software development environments (contexts).

\section{Discussion}

When selecting optimal defect prediction set-up for further verification if code smell-based metrics can improve prediction results, we observed that best result was achieved for dataset with bad smell metrics included (F-measure = 0.9713). However, for the same setup, but without code smells metrics, F-measure value was only by 0.0059 lower (Table~\ref{table:Results}) what makes the difference between SourceMonitor and CodeAnalysis results negligible. Final results collected from 20 different software sub-modules confirmed that statement: Average accuracy value for prediction based on dataset constructed basing on both sources was only by 0.0091 better than result for SourceMonitor-only based metrics (Average F-measure value difference = 0.0088), while standard deviation value was 0.0136. Worth noticing is drop of CodeAnalysis-only based prediction results, when feature selection (FS) process was removed from the experimental setup.

Results of our experiment of using code smells metrics in software defect prediction, show irrelevant -- in our opinion -- impact on effectiveness of the process, when basic dataset (SourceMonitor-based) was extended by CodeAnalysis metrics. because even if prediction effectiveness measures are slightly higher, the stay within the limits of error. But when only use of CodeAnalysis-based metrics were used for prediction (without basic set of SourceMonitor-based metrics), such process resulted with high accuracy (0.8249) and F-measure (0.8286) results.

Thus, answering the research question: \emph{How Code Bad Smells based metrics impact defect prediction in industrial software development project?} We want to state, that in industrial environment, such as PROSIT+ software development project, impact of code bad smells based metrics is negligibly small, and usage of CodeAnalysis-based metrics should not be considered useful, due to fact that additional effort needed for introducing code smell-based metrics to software defect prediction process is not compensated by relatively high increase of prediction effectiveness.

However, we observed surprisingly high effectiveness of prediction, when dataset based on CodeAnalysis only was used. Authors believe, that code bad smells can be effectively used for defect prediction process especially there, where other metrics are not available, or computing power is insufficient to handle large sets of different metrics (for example 24 kinds of metrics for SourceMonitor), while CodeAnalysis metrics set, used in our research, contained only 11 different kinds of metrics. Due these promising results, aspects of using code bad smells only based metrics in defect prediction processes should be investigated further.

\bibliographystyle{splncs03}
%\bibliography{bibliography}

\end{document}